\documentclass[twocolumn,showpacs,preprintnumbers,amsmath,amssymb,floatfix]{revtex4}

\usepackage{graphicx}
\usepackage{dcolumn}
\usepackage{bm}

\begin{document}

%

\let\a=\alpha      \let\b=\beta       \let\c=\chi        \let\d=\delta
\let\e=\varepsilon \let\f=\varphi     \let\g=\gamma      \let\h=\eta
\let\k=\kappa      \let\l=\lambda     \let\m=\mu
\let\o=\omega      \let\r=\varrho     \let\s=\sigma
\let\t=\tau        \let\th=\vartheta  \let\y=\upsilon    \let\x=\xi
\let\z=\zeta       \let\io=\iota      \let\vp=\varpi     \let\ro=\rho
\let\ph=\phi       \let\ep=\epsilon   \let\te=\theta
\let\n=\nu
\let\D=\Delta   \let\F=\Phi    \let\G=\Gamma  \let\L=\Lambda
\let\O=\Omega   \let\P=\Pi     \let\Ps=\Psi   \let\Si=\Sigma
\let\Th=\Theta  \let\X=\Xi     \let\Y=\Upsilon

%

%

\def\cA{{\cal A}}                \def\cB{{\cal B}}
\def\cC{{\cal C}}                \def\cD{{\cal D}}
\def\cE{{\cal E}}                \def\cF{{\cal F}}
\def\cG{{\cal G}}                \def\cH{{\cal H}}
\def\cI{{\cal I}}                \def\cJ{{\cal J}}
\def\cK{{\cal K}}                \def\cL{{\cal L}}
\def\cM{{\cal M}}                \def\cN{{\cal N}}
\def\cO{{\cal O}}                \def\cP{{\cal P}}
\def\cQ{{\cal Q}}                \def\cR{{\cal R}}
\def\cS{{\cal S}}                \def\cT{{\cal T}}
\def\cU{{\cal U}}                \def\cV{{\cal V}}
\def\cW{{\cal W}}                \def\cX{{\cal X}}
\def\cY{{\cal Y}}                \def\cZ{{\cal Z}}
%

\newcommand{\Ns}{N\hspace{-4.7mm}\not\hspace{2.7mm}}
\newcommand{\qs}{q\hspace{-3.7mm}\not\hspace{3.4mm}}
\newcommand{\ps}{p\hspace{-3.3mm}\not\hspace{1.2mm}}
\newcommand{\ks}{k\hspace{-3.3mm}\not\hspace{1.2mm}}
\newcommand{\des}{\partial\hspace{-4.mm}\not\hspace{2.5mm}}
\newcommand{\desco}{D\hspace{-4mm}\not\hspace{2mm}}



\title{\boldmath
 Remarks on the $A_{SL}^s$ and $\Delta\Gamma_s$ Studies at the Tevatron
 and Beyond
 }
\author{Wei-Shu Hou}
\author{Namit Mahajan}
\affiliation{
 Department of Physics, National Taiwan
 University, Taipei, Taiwan 10617, R.O.C.
}
\date{\today}

\begin{abstract}
We give an assessment of the three approaches, undertaken by the
D$\emptyset$ experiment, to probe $CP$ violation in $B_s$ mixing
without measurement of $B_s$-$\bar B_s$ oscillations or tagging:
dimuon charge asymmetry $A_{SL}$, untagged single muon charge
asymmetry $A_{SL}^s$, and lifetime difference in untagged $B_s \to
J/\psi\phi$ decay. The latter two approaches provide an
alternative avenue, if not crosscheck, to the usual mixing-decay
interference study in modes such as $B_s \to J/\psi\phi$.
Prospects at the Tevatron, LHC and (Super) B factories are
discussed.
\end{abstract}

\maketitle


$B_s$-$\bar B_s$ mixing has recently been measured by the CDF
experiment \cite{dmsCDF}. The result is consistent with the Standard
Model (SM) expectation, albeit slightly on the low side. This
suggests that New Physics (NP) corrections are mild, {\it unless}
the $CP$ violation (CPV) phase in $M_{12}$, $\phi_s \equiv
2\Phi_{B_s}$, is greatly different from SM expectation. Thus, the
focus has shifted to measurement of $\phi_s $. The standard approach
is via mixing-decay interference, or the so-called time-dependent
CPV studies, such as via the $B_s \to J/\psi\phi$ mode
\cite{dunietz}, which is analogous to $B_d \to J/\psi K_S$ studies
at the B factories. This approach depends on the ability of
resolving $B_s$-$\bar B_s$ oscillations, tagging the $B_s$ vs $\bar
B_s$ flavor, as well as having a large data sample of $B_s \to
J/\psi\phi$.

An alternative approach does not rely on measuring the very rapid
$B_s$-$\bar B_s$ oscillations, oftentimes forsaking even tagging.
The D$\emptyset$ experiment has recently conducted a three-pronged
approach along this line: the dimuon charge asymmetry $A_{SL}$
(self-tagged)~\cite{ASLD0}, the {\it untagged} single muon charge
asymmetry $A_{SL}^s$~\cite{ASLsD0}, and a measurement of the
lifetime difference in {\it untagged} $B_s \to J/\psi\phi$ decay,
but keeping the CPV phase $\phi_s$ free \cite{dgammaD0}. The three
measurements offer independent probes of CPV in the $B_s$ system,
although $A_{SL}$ also incorporates the effect from $B_d$ in the
Tevatron environment.
The experimental results, however, are not so easy to grasp. While
there are hints for large deviations from SM, at the same time
there are some conflicts.

In this note we comment on the merits and demerits of the three
different measurements, and assess future prospects at the
Tevatron, LHC, and the (Super)B factories.
Part of our motivation is whether one could have near maximal CPV in
$A_{SL}^s$. For this purpose, we take the 4 generation model
recently studied in \cite{4gen,4genbsmixing} as our NP scenario,
which we will show nearly saturates the possible enhancements from
SM in $A_{SL}^s$.

The same sign dimuon asymmetry
\begin{eqnarray}
A_{SL} &=& 4\,\frac{N_{b\bar b\to \ell^+\ell^+X}
             - N_{b\bar b\to \ell^-\ell^-X}}
              {N_{b\bar b\to \ell^+\ell^+X}
             + N_{b\bar b\to \ell^-\ell^-X}}, 
 \label{ASL}
\end{eqnarray}
is an experimental observable.
At the B factories running on the
$\Upsilon(4S)$ resonance, it directly gives $A_{SL} = A_{SL}^d$,
the so-called flavor-specific CPV asymmetry that measures CPV in
$B_d$ mixing alone. That is,
\begin{equation}
A_{SL}^{\Upsilon(4S)} = A_{SL}^d = 4\, \frac{{\rm
Re}\,\varepsilon_{B_d}}{1+\vert \varepsilon_{B_d}\vert^2},
\end{equation}
the quantity analogous to $\varepsilon_{K}$, the first observed
CPV effect in the kaon system.
The current values from the B factories are
\cite{asldbabar,asldbelle}
\begin{eqnarray}
A_{SL}^d &=& + 0.0016 \pm 0.0056 \pm 0.0040,
                         \,\;\ \ \ \ \ {\rm (BaBar)} \nonumber \\
         &=& - 0.0012 \pm 0.0080 \pm 0.0082.
                         \ \ \ \ \ \ \ {\rm (Belle)}
\end{eqnarray}
Note that the central values are opposite in sign, and both are
somewhat too consistent with zero compared to the allowed errors.

At hadronic colliders (and for B factories running above
$\Upsilon(4S)$ energies, e.g. on $\Upsilon(5S)$), the situation is
more complicated, as both $B_d$ and $B_s$ mesons are produced. But
this offers one the opportunity to measure $A_{SL}^s$. At the
Tevatron one has
\begin{eqnarray}
A_{SL}^{\rm TeV} &\simeq& 0.6 A_{SL}^d + 0.4 A_{SL}^s,
 \label{ASL}
\end{eqnarray}
where we follow the estimate of Ref.~\cite{nir} that considers the
production fractions of $B_d$ and $B_s$, and differences in the
evolution of these two systems.
The D$\emptyset$ experiment has recently measured~\cite{ASLD0}
\begin{equation}
A_{SL}^{\rm TeV} = -0.0092 \pm 0.0044 \pm 0.0032,
                         \ \ \ \ \ \ \ ({\rm D}\emptyset)
\end{equation}
where the superscript of TeV is ours, to distinguish from the
$\Upsilon(4S)$ environment of Eq. (2)

Making the naive world average of direct $A_{SL}^d$ measurements
of Eq. (3), including the result of CLEO \cite{cleo}, one obtains
$A_{SL}^d = +0.0011\pm 0.0055$. Combining with Eq.~(5), one gets
\begin{equation}
A_{SL}^s = -0.025 \pm 0.016.
                         \ \ \ \ \ \ \ \ \ \ \ (A_{SL}^d\ {\rm derived})
\end{equation}
The central value of Eq.~(6) is more negative than Eq.~(5) because
the current experimental average for $A_{SL}^d$ is positive. If
one uses the more conservative PDG 2006 average~\cite{PDG} of
$A_{SL}^d= -0.0052 \pm 0.0116$, which is negative, one gets
$A_{SL}^s = -0.015 \pm 0.022$, which is less negative than
Eq.~(6). In any case, Eq.~(6) is still consistent with zero. [With
a more detailed analysis than our approximate Eq.~(4) and a
different $A_{SL}^d$ world average, D$\emptyset$ gives the value
of $-0.0064\pm 0.0101$; see Note Added].

The measurement of the time integrated flavor {\it untagged}
charge asymmetry in semileptonic $B_s$ decay, the first of its
kind, gives \cite{ASLsD0}
\begin{eqnarray}
A_{SL}^{s,\,\rm untagged} &=& \frac{N_{B_s \to \mu^+D_s^-X}
                 - N_{B_s \to \mu^-D_s^+X}}
                  {N_{B_s \to \mu^+D_s^-X}
                 - N_{B_s \to \mu^-D_s^+X}}
      \nonumber \\
         \cong\; \frac{1}{2} A_{SL}^s &=& + 0.0123 \pm 0.0097 \pm 0.0017,
                         \ \ ({\rm D}\emptyset)
 \label{ASLs}
\end{eqnarray}
where the second step follows from $\Delta \Gamma_s^2 \ll \bar
\Gamma_s^2 \ll \Delta m_s^2$. Eq.~(7) is also consistent with
zero, but is at some variance with Eq.~(6): the central values are
larger than the errors, but have opposite sign. We wish to offer
some scrutiny of these measurements and their current and future
interpretation.

We deemphasize theoretical formalism, but it is useful to recall
that~\cite{dunietz}
\begin{eqnarray}
A_{SL}^s = {\rm Im}\, \frac{\Gamma_{12}^s}{M_{12}^s}
         = \left\vert \frac{\Gamma_{12}^s}{M_{12}^s} \right\vert \sin\phi_s
         = \frac{\Delta\Gamma_s}{\Delta m_s}\tan\phi_s.
\end{eqnarray}
In early times~\cite{hagelin}, various estimates of
$\vert\Gamma_{12}^s/M_{12}^s\vert$ were made, with large
uncertainties. Now, $\Delta m_s$ is well measured, whereas $\Delta
\Gamma_s$ can be extracted from data. It is fair to say that
$\Delta\Gamma_s/\Delta m_s \lesssim 10^{-2}$ from direct
measurement, hence both central values of Eqs.~(6) and (7) are
hard to sustain. To see this, it is instructive to turn to the
third measurement done by D$\emptyset$.

The finite width difference between the two $B_s$ states offers an
interesting probe of CPV. The $J/\psi\phi$ final state, being
$VV$, has $s$, $d$ wave and $p$ wave components, which are $CP$
even and odd, respectively. If $CP$ were conserved, the $CP$ even
and odd states would be the mass eigenstates. But with CPV, the
physical mass eigenstates have both $CP$ even and odd components,
hence they interfere in the time evolution of $B_s\to J/\psi\phi$
decay. The former implies that the $CP$ even (odd) state picks up
a $1-\cos\phi_s$ fraction of $CP$ odd (even) component, which is
completely analogous to the mixing of $CP$ eigenstates $K_{1,2}$
into the physical $K_{L,S}$ mesons. The $CP$ admixture leads to a
dilution of the lifetime difference~\cite{dunietz,grossman},
\begin{eqnarray}
\Delta\Gamma_s = \Delta\Gamma_s^{\rm SM} \, \cos\phi_s,
\end{eqnarray}
from the SM value $\Delta\Gamma_s^{\rm SM}$. With CPV in the time
evolution (i.e. $B_s$-$\bar B_s$ mixing and decay) of the physical
mass eigenstates, the $CP$ even and odd components interfere. That
is, CPV in mixing-decay interference leads to interference between
$s$, $d$ waves and $p$ wave components of opposite mass
eigenstates. This gives rise to a $\sin\phi_s$ term~\cite{dunietz}
in the differential rate, which has no analogue in the study of
$K_{L,S}\to \pi\pi$ decays, and in principle allows one to probe
the sign of $\phi_s$.

Doing an angular analysis of the time evolution of {\rm flavor
untagged} $B_s\to J/\psi\phi$ decays, D$\emptyset$ fits for width
difference $\Delta\Gamma_s$, mean width $\bar\Gamma_s$, and CPV
phase $\phi_s$, as well as the magnitudes and relative phases of
the decay amplitudes \cite{dgammaD0}, finding
\begin{equation}
\phi_s = -0.79 \pm 0.59,
                    \ \Delta\Gamma_s = 0.17 \pm 0.09\ {\rm ps}^{-1}.
                    \ \ ({\rm D}\emptyset)
\end{equation}
Fixing $\phi_s = 0$ in the fit, D$\emptyset$ finds $\Delta\Gamma_s
= 0.12^{+0.08}_{-0.10}$ ps$^{-1}$, which is fully consistent with
scaling $\Delta\Gamma_s = 0.17$ ps$^{-1}$ by $\cos\phi_s$ of
Eq.~(9). The value for $\Delta\Gamma_s$ in Eq.~(10) is smaller
than the values obtained previously by CDF and
D$\emptyset$~\cite{dgammas05} with smaller datasets, with $\phi_s$
held fixed at $0$.
Note that the current theory prediction gives $\Delta
\Gamma_s^{\rm SM} = 0.09 \pm 0.02$ ps$^{-1}$~\cite{dgammaSM}. The
one sigma range from Eq.~(10) gives $\vert\cos\phi_s\vert = 0.22$
to $0.98$, which {\it can support a finite $\phi_s$}. We will not
pursue the issue of sign ambiguities arising from relative sign
flips between angular amplitudes.

We can now return to the discussion of $A_{SL}^s$. Combining
Eqs.~(8) and (9), we see that
\begin{eqnarray}
\vert A_{SL}^s\vert
 \; < \; \frac{\Delta\Gamma_s^{\rm SM}}{\Delta m_s^{\rm CDF}}
 \; \lesssim \; 0.005,
\end{eqnarray}
if current theory prediction for $\Delta\Gamma_s^{\rm SM}$
\cite{dgammaSM} is to be trusted. This should make clear that the
central values of Eqs.~(6) and (7), i.e. $\mp 0.025$, are rather
unlikely. Put another way, if future measurements establish the
central values of Eqs.~(6) and (7) or analogous large values, then
$\Delta\Gamma_s^{\rm SM}$ is much larger than current theoretical
predictions, and likely has a rather nonperturbative, hadronic
enhancement that has to be understood. At present, since the two
approaches of Eqs.~(6) and (7) are basically independent, while
the signs of their central values are opposite hence compensate
each other, the combined result is still quite acceptable.

We now discuss the merits and demerits of the two approaches to
$A_{SL}^s$. Although the dimuon $A_{SL}^{\rm TeV}$ measurement,
Eq.~(5), has the advantage of statistics compared to the single
muon $A_{SL}^{s,\,\rm untagged}$ measurement, Eq.~(7), it is
indirect and already close to being systematic limited. The
requirement of positive muon identification for both muons makes
it sensitive~\cite{ASLD0} to a kaon charge asymmetry that is
induced by $K^-$ interaction with detector material. This brings
in a bias for muons entering the muon detector (farthest from
interaction point) towards $K^+\to \mu^+\nu$ decay. The central
value of Eq.~(5) is in fact mostly due to~\cite{ASLD0} a
correction from the measurement of this matter-induced charge
asymmetry from $K^\pm \to \mu^\pm\nu$ decay muons , which also
dominates the systematic error of $A_{SL}^{\rm TeV}$ measurement.
We see that the extraction of $A_{SL}^s$ from dimuon asymmetry,
$A_{SL}^{\rm TeV}$, would be harder to improve, which is further
compounded by the need to disentangle the contribution from
$A_{SL}^d$. Note that the fractional contribution of $A_{SL}^d$ vs
$A_{SL}^s$ itself in Eq.~(4) carry errors.

The single muon charge asymmetry $A_{SL}^{s,\,\rm untagged}$
recently studied also by D$\emptyset$~\cite{ASLsD0} is, in
contrast, rather interesting. The systematic error of Eq.~(7) for
$A_{SL}^s = 2A_{SL}^{s,\,\rm untagged}$ is 0.0035, which is less
than Eq.~(11), and one is limited by the statistical error of
0.0193 at present. The event signature is $B_s \to \mu^\pm
D_s^\mp+X$, which suppresses contamination from $B_d$. One
reconstructs $D_s^\pm \to \phi\pi^\pm$, with $\phi\to K^+K^-$,
which is insensitive to kaon charge asymmetry. The $\pi^\pm$
charge identification is from the tracking system, hence suffers
much less material effect. To further cancel the charge asymmetry
for the $\mu^\pm$ and the $\pi^\mp$, D$\emptyset$ periodically
reverses~\cite{ASLD0,ASLsD0} the polarity of their magnet. Thus,
although $D_s$ reconstruction reduces the number of signal events,
the gain in systematic control is significant.
However, even if the systematic error can be improved with
increase of statistics, one would still need more than
15~fb$^{-1}$ of data. Otherwise, a factor of two or more
improvement in efficiency (e.g. by studying other modes) with 8
fb$^{-1}$ data is needed, without bringing in further systematic
errors, for one to reach the sensitivity of Eq.~(11). So the
situation seems challenging at the Tevatron. The method could be
interesting for the LHC, but one would have to deal with
production bias since one has $pp$ rather than $p\bar p$
production environment.

Eq.~(11) is only a bound. To motivate further pursuit, we ask: Can
it be saturated in reality? The answer is in the affirmative, and
we give a realistic example as existence proof.

A sequential 4th generation (SM4) may be called for~\cite{4gen}
from the time-dependent and direct CPV ``anomalies"~\cite{HFAG} of
$\Delta {\cal S} \equiv S_{b\to s\bar qq} - S_{b\to s\bar cc} < 0$
(2.6 $\sigma$) and $\Delta {\cal A}_{K\pi} \equiv {\cal
A}_{K^+\pi^-} - {\cal A}_{K^+\pi^0} < 0$ ($> 4\sigma$). The recent
measurement of $B_s$ mixing, together with $B\to X_s\ell^+\ell^-$
rate, although consistent with SM, could alternatively
imply~\cite{4genbsmixing} that $\sin2\Phi_{B_s} = \sin\phi_s \sim
-0.4$ to $-0.7$ in SM4, with typical value at $-0.55$. This give
$\cos\phi_s \sim 0.84$, or
\begin{equation}
\phi_s^{\rm SM4} \sim -0.6,
\end{equation}
which is consistent with Eq.~(10). With $\tan\phi_s \sim -0.66$,
using $\Delta\Gamma_s$ from Eq.~(10) in Eq.~(8), one basically
saturates the bound of Eq.~(11).
$\Delta\Gamma_s$ may be lower than Eq.~(9), but a larger value of
$\sin\phi_s$ is certainly allowed by varying $f_{B_s}^2 B_{B_s}$
(and $m_{t'}$).
We stress that the current B factory result of $\Delta {\cal S}
\lesssim 0$, as well as $\Delta {\cal A} > 0$, though still
inconclusive, give prime motivation for study of CPV in $B_s$
oscillations and decay, with the 4th generation model just an
example.

Note that a larger $|\sin\phi_s|$ means a smaller $|\cos\phi_s|$,
which would suppress $\Delta\Gamma_s$ measured in flavor-specific
$B_s$ decays. Taking the ratio with $\Delta\Gamma_s$ measured in
$CP$ eigenstate, such as $B_s\to D_s^+D_s^-$, can give a measure
of $|\cos\phi_s|$~\cite{dunietz,grossman},
\begin{equation}
\frac{\Delta\Gamma_s}{\Delta\Gamma_s^{CP}} = \cos\phi_s.
\end{equation}
This is complementary to $\sin\phi_s$ measurement. The value
predicted by the 4th generation model is 0.7 to 0.9.

The lifetime difference and $A_{SL}^s$ approach allows a potential
dark horse in the race: B factories. Recently, the Belle
experiment has taken over 20 fb$^{-1}$ data on $\Upsilon(5S)$, and
has confirmed~\cite{Drutskoy} the $B_s$ signal already seen by the
CLEO experiment~\cite{CLEO}.
$B_s$ events can be kinematically separated from $B_d$, but time
dependent CPV in $B_s$ decays is clearly out of reach. Because of
a low cross section for $e^+e^- \to \Upsilon(5S)$ and a low
fraction of $\Upsilon(5S)$ branching into $B_s^{(*)}\bar
B_s^{(*)}$ ($\sim 18\%$~\cite{Drutskoy,CLEO}), the 20 fb$^{-1}$
data at hand contains only several million $B_s$ mesons. Factoring
in decay branching ratios and various detection efficiencies, one
at best can reach a statistical error of a few percent for
$A_{SL}^{s,\,\rm untagged}$. To measure the lifetime difference,
whether or not in $CP$ eigenstates, one suffers from similar low
effective production and detection efficiency. On the other hand,
it seems unlikely for B factories to run on the $\Upsilon(5S)$ for
several hundred fb$^{-1}$ before the arrival of the Super B
factory era. It thus seems that the B factories would not be able
to compete with the Tevatron on $A_{SL}^{s,\,\rm untagged}$ and
$\Delta\Gamma_s^{(CP)}$. However, by studying~\cite{druts_moriond}
the branching ratios of decay modes such as $B_s\to D_s^{(*)} \bar
D_s^{(*)}$, which may be at the root of $\Delta\Gamma_s \neq 0$,
the B factory can definitely shed further light on the issue, the
more data on the $\Upsilon(5S)$ the better. $A_{SL}^s$ and
$\Delta\Gamma_s$ can certainly be studied easily at a Super B
factory with a couple orders of magnitude improvement in
luminosity. In that era, one would have to compete with detectors
like LHCb that already has ran for a few years.

We remark that $A_{SL}^d$ has to be SM-like, given that
$\sin\phi_d \equiv \sin2\Phi_{B_d}$ is found to be fully
consistent with SM. This applies also to SM4, and it has been
demonstrated~\cite{phid} to be the case even with large
$|\sin\phi_s|$, once the kaon constraints are applied. Thus,
$A_{SL}^d$ is also out of reach at present B factories, and would
be Super B factory physics. Judging from the errors in Eq.~(3),
this would require meticulous control of systematic errors.

In summary, finding $\sin\phi_s < -0.5$ as predicted e.g. by 4th
generation model would certainly be rather astounding. This is
more directly probed via mixing-decay interference studies, e.g.
in $B_s \to J/\psi\phi$ mode pursued by both CDF and D$\emptyset$,
and can be perfected by LHCb. However, D$\emptyset$ has
demonstrated an alternative approach. The angular resolved time
evolution of {\rm flavor untagged} $B_s\to J/\psi\phi$ decays
probes $\Delta\Gamma_s$, as well as both $\cos\phi_s$ and
$\sin\phi_s$. The single muon {\it untagged} charge asymmetry
$A_{SL}^{s,\,\rm untagged}$ in semileptonic $B_s$ decay probes the
product of $\Delta\Gamma_s$ and $\tan\phi_s$. Assisted by the B
factory studies on $\Upsilon(5S)$, it is not impossible that the
first hint for CPV in $B_s$ system may still come from, say,
$|\cos\phi_s| < 1$ from the angular resolved lifetime difference
study. Much depends on the sensitivity reach for $\sin\phi_s
\equiv \sin2\Phi_{B_s}$ via oscillation dependent CPV studies, and
the associated analysis time frame.

\vskip 0.3cm \noindent {\bf Acknowledgement}.\
 We thank B.~Casey, K.F. Chen, A.~Drutskoy, and T.~Nakada
 for discussions.
 This work is supported in part by NSC 95-2119-M-002-037 and NSC 95-2811-M-002-031
 of Taiwan.

\vskip 0.3cm \noindent {\bf Note Added}.\
 After our work became public, the D$\emptyset$ experiment posted the
 paper hep-ex/0702030, which combines the results of
 Refs.~\cite{ASLD0,ASLsD0,dgammaD0} together with the world average
 flavor specific $B_s$ lifetime. The approach to Eq.~(4) is more specific
 than the approximate values of Ref.~\cite{nir}, which we have taken.
 Second, the value for $A_{SL}^d$ used is somewhat different than,
 but consistent with, the two values that we have illustrated with
 (in and below our Eq.~(6)). The combined end result of
 D$\emptyset$ is $\Delta\Gamma_s = 0.13 \pm 0.09\ {\rm ps}^{-1}$ and
 $\phi_s = -0.70^{+0.47}_{-0.39}$. The former is more consistent with SM,
 while for the latter the errors have been reduced slightly, and,
 although only 1.2$\,\sigma$ from zero, makes our Eq.~(12) more intriguing.

\end{document}